\documentclass[floatfix,twocolumn,aps,prd,showpacs,amsmath,amssymb]{revtex4}
\usepackage{graphicx}
\usepackage{epsfig}
\usepackage{psfrag}
\newcommand{\be}{\begin{equation}}
\newcommand{\ee}{\end{equation}}

\begin{document}
%
\title{A principal component analysis for LISA -- the TDI connection}
%
%
\author{J. D. Romano$^{1}$\footnote{joseph.romano@astro.cf.ac.uk} and
G. Woan$^{2}$\footnote{graham@astro.gla.ac.uk}}
\affiliation{$^1$School of Physics and Astronomy, Cardiff
University, Cardiff, CF24\,3YB, UK\\
$^2$Department of Physics and Astronomy, University of Glasgow,
Glasgow, G12\,8QQ, UK}
\date{\today}
%
\begin{abstract}
Data from the Laser Interferometer Space Antenna (LISA) is expected
to be dominated by frequency noise from its lasers. However the
noise from any one laser appears more than once in the data and
there are combinations of the data that are insensitive to this
noise.  These combinations, called \emph{time delay interferometry
(TDI) variables}, have received careful study, and point the way to
how LISA data analysis may be performed.  Here we approach the
problem from the direction of statistical inference, and show that
these variables are a direct consequence of a principal component
analysis of the problem. We present a formal analysis for a simple
LISA model and show that there are eigenvectors of the noise
covariance matrix that do not depend on laser frequency noise.
Importantly, these orthogonal basis vectors correspond to linear
combinations of TDI variables. As a result we show that the
likelihood function for source parameters using LISA data can be
based on TDI combinations of the data without loss of information.
\end{abstract}
%
%
\maketitle
\section{Introduction}
The Laser Interferometer Space Antenna (LISA) is a space-borne
gravitational telescope currently under development by ESA and NASA
\cite{LISA}. It is designed to detect gravitational radiation at
frequencies between $\sim 10^{-5}$\,Hz and  $\sim 1$\,Hz, a band
that is not readily accessible from Earth due to local noise. The
current design comprises a constellation of three spacecraft in
circular orbits of radius 1\,AU around the Sun, in a
near-equilateral configuration of side $5\times10^6$\,km.
Gravitational waves passing through the telescope modulate the
separation between the spacecraft on scales of picometres.  This
modulation is sensed by laser beams exchanged between spacecraft,
and recorded as the difference in frequency between the
locally-generated and received laser signals. In simple terms there
are therefore three independent lasers in the system, and six raw
signal streams (referred to as \emph{Doppler measurements} in the
literature, e.g., \cite{livrev}) corresponding to the six
bidirectional baseline combinations of these lasers.

The main source of noise in these raw streams comes from the
relative frequency stability of the reference lasers, giving a
spectral density of $\sim 10^{-13}$\,Hz$^{-1/2}$ in the millihertz
band. This laser frequency noise results in a strain noise floor
$\sim 10^4$ times higher than the target sensitivity of LISA and, at
first sight, severely limits LISA's performance.  The scenario is
very similar to that encountered in radio astronomical very long
baseline interferometry (VLBI), where free-running local oscillators
are used at each end of the interferometric baseline, giving the
resultant fringes an unknown and varying fringe rate.  The
breakthrough in radio astronomy came with the realisation that with
three or more telescopes, and therefore three or more baselines, one
can form \emph{closure} quantities that are insensitive to the
relative phases of the local oscillators.  The simplest of these
relations is \emph{closure phase}, comprising the sum of
interferometric phases around a loop of baselines \cite{tms}.  The
phase of each local oscillator appears twice in such a sum and with
opposite signs, so that closure phase contains only
baseline-dependent (and therefore astronomical), rather than
antenna-dependent contributions. This direct method was used in the
early days of radio interferometry \cite{jennison}, in VLBI
\cite{rogers} and more recently in optical interferometry
\cite{coast}. In modern radio astronomy these principles are
encapsulated in the idea of \emph{self calibration}, in which each
antenna in the interferometer is allocated an unknown complex gain
factor \cite{cornwell, schwab}.  The interferometer data are then
used to simultaneously determine both these gain factors and the sky
map, so greatly increasing the performance of the instrument.

The situation for LISA is more complicated. Laser noise can be
canceled only if the closure relations take account of the light
travel time between spacecraft, and the presence of bidirectional
beams along the baselines increases the number of possible
relations.  This has led to the development of Time-Delay
Interferometry (TDI) variables for LISA \cite{tintoarm}.  These are
linear combinations of the six LISA data streams, suitably offset in
time to cancel the noise contributions from the three lasers.  TDI
variables have received detailed attention in the literature and
there is now a sophisticated understanding of their generation and
properties \cite{firstsens, dhurandhar, cornhell, shaddock2004,
geoTDI, livrev}, extending to `second-generation' TDI variables that
take account of the slight relative motion of the spacecraft
\cite{stea2003, tea2004}.

The very existence of TDI variables clearly shows that LISA is
capable of generating data that is sensitive to astronomical sources
but not to laser noise.  This is an important step, but it does not
tell us how to use these TDI variables to do astronomy.  For this we
need a method of using these derived data to constrain the sky and
make statements about the parameters of individual sources of
gravitational radiation.  The natural framework for such statements
is that of statistical inference, in which we construct a Bayesian
probability for a source parameter, or set of parameters, $a$, given
the LISA data, $d$.  Formally
 \be
 p(a|d)\propto p(a)\, p(d|a)\,,
 \label{e:bayes}
 \ee
where $p(d|a)$ is the probability of getting a certain set of data
given a noise model and a particular value for $a$. This is familiar
as Bayes' Theorem, and requires the prior probability for $a$,
$p(a)$, before it can be fully applied. Although priors form an
important part of any analysis we will not concentrate on them here
and they do not depend on the current data. The quantity $p(d|a)$ is
usually called the \emph{likelihood} of $a$ and, importantly, fully
defines how the data enters into the calculation. Indeed the
likelihood contains all that the data has to say on the matter, so
that the heart of a parameter estimation problem is fully defined
once a likelihood is written down.

We will show below that the likelihood function for LISA contains
several insights into the meaning and role of TDI variables. We
consider the noise covariance matrix for a series of LISA Doppler
measurements over several time steps. For Gaussian noise, the
inverse of this covariance matrix defines the log likelihood of the
parameters, and its quadratic form defines contours of equal
likelihood, in a space equal in dimension to the number of data
points used. The eigenvectors of this covariance matrix are the
principal axes of the equal-likelihood hyperellipsoidal surface, and
correspond to linear combinations of the data that give maximal and
minimal covariance. Principal component analysis (PCA) is simply the
process of identifying these eigenvectors and using only a subset of
them to characterise the data.  Usually it is the components with
the largest eigenvalues that are desired, as these are the data
combinations that contain the majority of the covariance. These
components are used in fields such as pattern recognition in order
to generalise or compress data sets. For LISA however, these
principal components are the ones that contain the highly correlated
laser frequency noise. In contrast therefore, we are interested in
the existence of eigenvalues that do \emph{not} depend on the laser
noise and are minimal.  We will show that the eigenvalues of the
LISA covariance matrix fall into two distinct groups, distinguished
by their dependence on laser noise. The group that is independent of
the common frequency noise corresponds directly to the TDI variables
considered above, in fact the relationship is hinted at in
\cite{livrev}. Not all will necessarily be sensitive to
gravitational wave signals, but as a group they orthogonally span
the data sub-space that corresponds to LISA's design noise floor.
The other group still contain astronomy, but are dominated by the
effects of laser frequency fluctuations.

In Section II we will consider a simple example of PCA to
highlight the essence of the method and demonstrate its
applicability to LISA data analysis.  In Section~III we extend
these ideas to tackle a real, but somewhat simplified, LISA model
and show that minor principal components are TDI variables.  In
Section~IV we discuss some of the immediate implications of this
work for the design and data analysis of LISA.

\section{Simple example}

Consider a single sample of data from two generic detectors
\begin{eqnarray}
s_1 &=& p + n_1 + h_1\,,
\\
s_2 &=& p + n_2 + h_2\,,
\end{eqnarray}
where $n_1$, $n_2$ are uncorrelated noises in the individual
detectors, $p$ is a common noise term, and $h_1$, $h_2$ are the
astrophysical signals of interest. For simplicity, we assume that
the noises are Gaussian-distributed with zero mean and variances
\begin{equation}
\langle n_1^2\rangle =
\langle n_2^2\rangle \equiv \sigma_n^2
\quad{\rm and}\quad
\langle p^2\rangle \equiv \sigma_p^2\,,
\label{e:variances}
\end{equation}
and that they are mutually uncorrelated, so that
\begin{equation}
\langle n_1 n_2\rangle =
\langle n_1 p\rangle =
\langle n_2 p\rangle
= 0\,.
\label{e:covariances}
\end{equation}
In addition, we assume a simple model in which the astrophysical
signals in the two detectors are
\begin{equation}
h_1 = 2a
\quad{\rm and}\quad
h_2 = a\,,
\end{equation}
where $a$ is a fixed but unknown constant whose posterior
distribution, $p(a|s_1,s_2)$, we want to compute.

To do this, in accordance with Eq.~(\ref{e:bayes}), we need to
calculate the likelihood $p(s_1,s_2|a)$. For Gaussian noise,
\begin{equation}
p(s_1,s_2|a) \propto \exp\left[-\frac{1}{2}Q\right]\,,
\end{equation}
where
\begin{eqnarray}
Q
&\equiv&
({\bf s}-{\bf h})^{\rm T}\cdot C^{-1}\cdot ({\bf s}-{\bf h})
\\
&\equiv&
\sum_{i,j=1}^2 (s_i-h_i)C^{-1}{}_{ij}(s_j-h_j)
\end{eqnarray}
is a quadratic form, involving the inverse of the noise
covariance matrix $C$ whose elements are
\begin{equation}
C_{ij}\equiv\langle (s_i-h_i)(s_j-h_j)\rangle\,.
\end{equation}
Using Eqs.~(\ref{e:variances}-\ref{e:covariances}), we find
\begin{equation}
C=
\left(
\begin{array}{cc}
\sigma_p^2+\sigma_n^2 &
\sigma_p^2 \\
\sigma_p^2 &
\sigma_p^2+\sigma_n^2 \\
\end{array}
\right)\,.
\label{e:C}
\end{equation}

Principal component analysis simplifies the calculation of the
likelihood $p(s_1,s_2|a)$, by identifying the eigenvectors of this
covariance matrix $C$. Note that the eigenvectors of $C$ are also
the eigenvectors of $C^{-1}$, but with reciprocal eigenvalues.  The
eigenvectors of $C$ in Eq.~(\ref{e:C}) are
\begin{equation}
{\bf e}_+ \equiv
\frac{1}{\sqrt{2}}
\left(
\begin{array}{c}
1\\
1\\
\end{array}
\right)
\quad{\rm and}\quad
{\bf e}_- \equiv
\frac{1}{\sqrt{2}}
\left(
\begin{array}{c}
1\\
-1\\
\end{array}
\right)\,,
\end{equation}
with eigenvalues
\begin{equation}
\lambda_+ \equiv 2\sigma_p^2 +\sigma_n^2 \quad{\rm and}\quad
\lambda_- \equiv \sigma_n^2
\end{equation}
respectively.
If we form the matrix of eigenvectors
\begin{equation}
E\equiv
\left(
\begin{array}{cc}
{\bf e}_+ & {\bf e}_-\\
\end{array}
\right)
=
\frac{1}{\sqrt{2}}
\left(
\begin{array}{cc}
1 & 1\\
1 & -1\\
\end{array}
\right)
\end{equation}
and use the facts that
\begin{eqnarray}
&&E\cdot E^T = E^T\cdot E = {\rm I}\,,
\\
&&E\cdot C^{-1}\cdot E^T
=\left(
\begin{array}{cc}
1/\lambda_+ & 0\\
0 & 1/\lambda_-\\
\end{array}
\right)
\equiv \Lambda^{-1}\,,
\end{eqnarray}
and
\begin{equation}
E\cdot ({\bf s}-{\bf h}) =
\frac{1}{\sqrt{2}}
\left(
\begin{array}{c}
(s_1+s_2)-(h_1+h_2)\\
(s_1-s_2)-(h_1-h_2)\\
\end{array}
\right)\,,
\end{equation}
we can rewrite $Q$ as
\begin{align}
Q &=({\bf s}-{\bf h})^{\rm T}\cdot C^{-1}\cdot ({\bf s}-{\bf h})
\\
&=({\bf s}-{\bf h})^{\rm T}\cdot (E^T\cdot E)\cdot C^{-1}\cdot
(E^T\cdot E)\cdot ({\bf s}-{\bf h})
\\
&=\left(E\cdot ({\bf s}-{\bf h})\right)^T\cdot \Lambda^{-1}
\cdot \left(E\cdot ({\bf s}-{\bf h})\right)\\
&= \frac{1}{2}\frac{1}{2\sigma_p^2+\sigma_n^2}\left(s_+-3a\right)^2+
\frac{1}{2}\frac{1}{\sigma_n^2}\left(s_--a\right)^2\,,
\end{align}
where we have defined $s_+$, $s_-$ to be the the eigencombinations
\begin{eqnarray}
s_+&\equiv& s_1+s_2\,,
\label{e:s_+}
\\
s_-&\equiv& s_1-s_2\,.
 \label{e:s_-}
\end{eqnarray}
Thus, the likelihood $p(s_1,s_2|a)$ factorises into a product
of likelihoods
\begin{equation}
p(s_1,s_2|a)\propto
p(s_+|a)p(s_-|a)\,,
\label{e:factorise}
\end{equation}
where
\begin{eqnarray}
p(s_+|a)&\propto&
\exp\left[-\frac{1}{2}
\frac{(s_+-3a)^2}{4\sigma_p^2+2\sigma_n^2}\right]\,,
\label{e:p(s_+|a)}
\\
p(s_-|a)&\propto&
\exp\left[-\frac{1}{2}
\frac{(s_--a)^2}{2\sigma_n^2}\right]\,.
\label{e:p(s_-|a)}
\end{eqnarray}

\begin{figure}[htbp!]
\includegraphics[width=2.5in,angle=0]{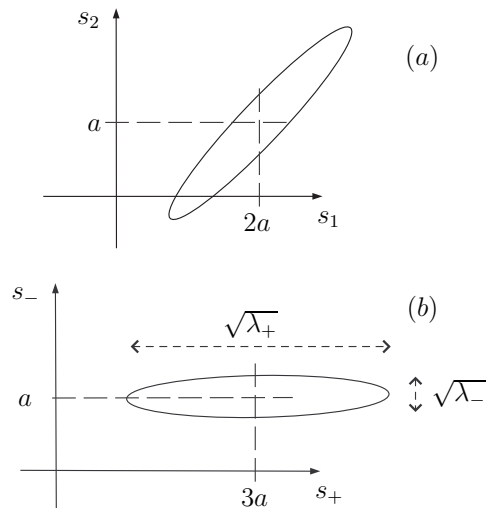}
\caption{Data probability contours referred to $(a)$ the original
data axes and $(b)$ axes corresponding to the eigen-combinations of
the data (with zero covariance). } \label{fig:cov}
\end{figure}

Note that the $s_-$ data combination, Eq.~(\ref{e:s_-}),
corresponds to the {\em minimum} eigenvalue $\lambda_-
\equiv\sigma_n^2$, which is independent of the common noise
variance $\sigma_p^2$. In this sense, $s_-$ is the {\em preferred}
data combination for this example, since it is least affected by
the various noise terms (see Fig.~\ref{fig:cov}).
Moreover, if $\sigma_p^2\gg\sigma_n^2$, as is the case for the
common laser frequency noise for LISA, then {\em there is
(effectively) no loss in information by doing statistical
inference with only the $s_-$ data combination}. This is most
easily seen by writing down the posterior distribution for $a$:
\begin{eqnarray}
p(a|s_1,s_2) &\propto& p(s_1,s_2|a)\,p(a)
\\
&\propto& p(s_+|a)\,p(s_-|a)\,p(a)
\\
&\propto& p(s_-|a)\,p(a)\,,
\end{eqnarray}
where the last proportionality follows from the fact that the
Gaussian $p(s_+|a)$ is effectively constant over the range of
$a$-values for which $p(s_-|a)$ is peaked (cf.\
Eqs.~(\ref{e:p(s_+|a)}-\ref{e:p(s_-|a)}) with $\sigma_p^2\gg
\sigma_n^2$). Thus PCA has simplified the analysis of this
particular problem by identifying a combination of the original data
(in this case $s_-\equiv s_1-s_2$) that captures nearly all the
available information on our parameter $a$.

\section{LISA example}

\begin{figure}[htbp!]
\includegraphics[width=2.2in,angle=0]{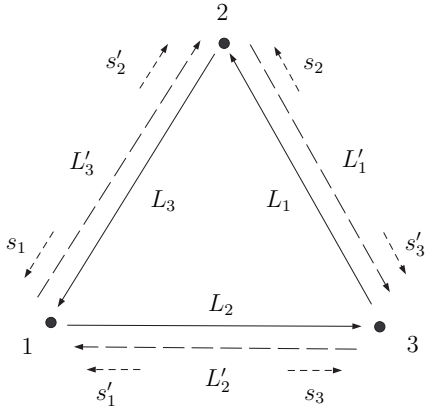}
\caption{Schematic of LISA configuration, following the conventions
in \cite{livrev}. The spacecraft are labeled 1, 2, 3. The separation
between spacecraft are denoted by $L_i$, $L'_i$, where the index $i$
corresponds to the opposite spacecraft. The beam arriving at
spacecraft $i$ has subscript $i$ and is unprimed or primed depending
on whether the beam is traveling counter-clockwise or clockwise
around the LISA triangle, as viewed from above.}
\label{fig:triangle}
\end{figure}

As mentioned in Sec.~I, the current design of LISA comprises three
spacecraft in circular orbits of radius 1~AU around the Sun, in a
near equilateral configuration of side $5\times 10^6$. The tiny
(picometre) modulation of the separation between the spacecraft
produced by the passage of gravitational waves is sensed by Doppler
measurements of laser beams exchanged between spacecraft, recorded
as the difference in frequency between the locally-generated and
received laser signals at each spacecraft. Since each spacecraft
receives laser signals sent from the other two spacecraft, there are
six raw data streams in total, denoted $s_1$, $s'_1$, $s_2$, $s'_2$,
$s_3$, $s'_3$, where the subscript indicates the spacecraft
receiving the laser beam, and is unprimed or primed depending on
whether the beam is traveling counter-clockwise or clockwise around
the LISA triangle, as viewed from above (see
Fig.~\ref{fig:triangle}).

Following the notation in \cite{livrev}, we write the six
LISA data streams as
\begin{eqnarray}
s_1  &=&{\cal D}_3p_2-p_1 + n_1 + h_1\,,
\\
s'_1 &=&{\cal D}_2p_3-p_1 + n'_1 + h'_1\,,
\end{eqnarray}
together with their cyclic permutations ($1\rightarrow 2\rightarrow
3\rightarrow 1$) at spacecrafts 2 and 3. Here $p_i$ is the frequency
noise associated with the two lasers (assumed for now to be locked
to one another) on spacecraft $i$ ($i=1,2,3$); $n_i$ and $n'_i$ are
all other noises associated with the transmission of the signal to
spacecraft $i$ in the counter-clockwise and clockwise directions,
and $h_i$ and $h'_i$ are the frequency modulations produced by the
astrophysical signals. ${\cal D}_i$ is a delay operator that takes a
data stream $x(t)$ and delays it by the light travel time down the
arm $L_i$:
\begin{equation}
{\cal D}_i x(t) = x(t-L_i)\,,
\end{equation}
in units where the speed of light $c=1$.
Explicitly,
\begin{eqnarray}
s_1(t)  &=& p_2(t-L_3)-p_1(t) + n_1(t) + h_1(t)\,,
\label{e:s1(t)}
\\
s'_1(t) &=& p_3(t-L_2)-p_1(t) + n'_1(t) + h'_1(t)\,,
\label{e:s1'(t)}
\end{eqnarray}
and similarly for the data streams at spacecrafts 2 and 3.
Note that we are restricting ourselves in this example to
the case of a non-rotating LISA configuration with fixed
arm-lengths.
We assume that the light travel time down an arm is
independent of the direction in which it is moving
(counter-clockwise or clockwise) and is independent of the
time of emission.
This corresponds to $L_i=L'_i$ in Fig.~\ref{fig:triangle}.

In practice, the data streams will be discretely-sampled
on-board the spacecrafts, with sampling period $\Delta t$.
For simplicity, we assume that the light travel times
down the arms are related by simple integers---in particular,
\begin{equation}
L_1 = \Delta t\,,
\quad
L_2 = 2\,\Delta t\,,
\quad
L_3 = 3\,\Delta t\,.
\label{e:armlengths}
\end{equation}
This restriction is made only to minimise the number of data
points needed to illustrate the PCA method.
It should be relatively straightforward to extend our analysis
to the case of more complicated light travel times, as well as
to situations where there is relative motion between the spacecraft
(i.e., $L_i\ne L_i'$).

If we denote the discrete time stamps by
$t_\alpha \equiv \alpha\,\Delta t$
and the value of data stream $x(t)$ at $t=t_\alpha$ by
\begin{equation}
x[\alpha]\equiv x(t_\alpha)\,,
\end{equation}
then the values of the six LISA data streams at time stamp
$\alpha=1$ become
\begin{eqnarray}
s_1[1] &=& p_2[-2]-p_1[1]+n_1[1]+h_1[1]\,,
\\
s'_1[1] &=& p_3[-1]-p_1[1]+n'_1[1]+h'_1[1]\,,
\\
s_2[1] &=& p_3[0]-p_2[1]+n_2[1]+h_2[1]\,,
\\
s'_2[1] &=& p_1[-2]-p_2[1]+n'_2[1]+h'_2[1]\,,
\\
s_3[1] &=& p_1[-1]-p_3[1]+n_3[1]+h_3[1]\,,
\\
s'_3[1] &=& p_2[0]-p_3[1]+n'_3[1]+h'_3[1]\,,
\end{eqnarray}
where we used Eq.~(\ref{e:armlengths}) to explicitly
evaluate the arguments of the time-delayed laser frequency noise.
To obtain expressions for the data streams evaluated at
other time stamps, we simply increment or decrement the
arguments of the data streams by the appropriate number
of sampling periods.

To simplify the calculation further, we will only consider data
streams having time stamps $\alpha=1,2,3,4,5$. Since there are six
data streams in total, this corresponds to a 30-dimensional vector
space of data points. The noise covariance $C$ is thus a $30\times
30$ matrix, whose $\alpha\beta$th element is itself a $6\times 6$
matrix:
\begin{widetext}
\begin{equation}
C_{\alpha\beta} \equiv
\left(
\begin{array}{cccccc}
\langle(s_1[\alpha]-h_1[\alpha])(s_1[\beta]-h_1[\beta])\rangle &
\langle(s_1[\alpha]-h_1[\alpha])(s'_1[\beta]-h'_1[\beta])\rangle &
{\rm etc.} & \cdot & \cdot & \cdot \\
\langle(s'_1[\alpha]-h'_1[\alpha])(s_1[\beta]-h_1[\beta])\rangle &
\langle(s'_1[\alpha]-h'_1[\alpha])(s'_1[\beta]-h'_1[\beta])\rangle &
\cdot & \cdot & \cdot & \cdot \\
\langle(s_2[\alpha]-h_2[\alpha])(s_1[\beta]-h_1[\beta])\rangle &
{\rm etc.} & \cdot & \cdot & \cdot & \cdot \\
\langle(s'_2[\alpha]-h'_2[\alpha])(s_1[\beta]-h_1[\beta])\rangle &
\cdot & \cdot & \cdot & \cdot & \cdot \\
\langle(s_3[\alpha]-h_3[\alpha])(s_1[\beta]-h_1[\beta])\rangle &
\cdot & \cdot & \cdot & \cdot & \cdot \\
\langle(s'_3[\alpha]-h'_3[\alpha])(s_1[\beta]-h_1[\beta])\rangle &
\cdot & \cdot & \cdot & \cdot & \cdot \\
\end{array}
\right)\,.
\end{equation}
\end{widetext}
As the matrix $C_{\alpha\beta}$ depends only on the difference
between $\alpha$ and $\beta$, (i.e., $C_{11}=C_{22}$,
$C_{12}=C_{23}$, etc.), the full covariance matrix $C$ is
\emph{block Toeplitz} (i.e., it is constant along the
$\alpha$-$\beta$ diagonals). Since $C$ is also symmetric, we need
only calculate $C_{11}$, $C_{12}$, $C_{13}$, $C_{14}$, and $C_{15}$
to fully determine $C$.

\begin{figure}[htbp!]
\includegraphics[width=2in,angle=0]{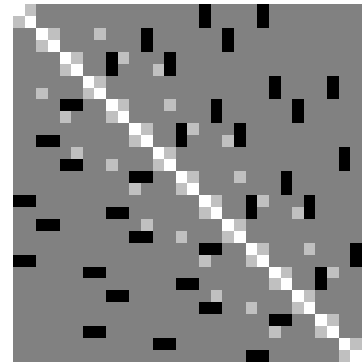}
\caption{Schematic representation of the covariance matrix, $C$, for
the simple LISA model. Blocks of increasing grey represent values
of $2\sigma_p^2 +\sigma_n^2$, $\sigma_p^2$, $0$  and $-\sigma_p^2$.}
\label{fig:cov_schematic}
\end{figure}

Finally, we assume that the laser frequency noise $p_i[\alpha]$
and individual noise terms $n_i[\alpha]$, $n'_i[\alpha]$ are
Gaussian-distributed with zero-mean and variances
\begin{eqnarray}
&&\langle n_i[\alpha]n_j[\beta]\rangle =
\langle n'_i[\alpha]n'_j[\beta]\rangle =
\delta_{\alpha\beta}\delta_{ij}\sigma_n^2\,,
\\
&&\langle p_i[\alpha]p_j[\beta]\rangle =
\delta_{\alpha\beta}\delta_{ij}\sigma_p^2\,,
\end{eqnarray}
(i.e., the random processes are white), and that they are
mutually uncorrelated, so that
\begin{equation}
\langle n_i[\alpha]n'_j[\beta]\rangle=
\langle n_i[\alpha]p_j[\beta]\rangle=
\langle n'_i[\alpha]p_j[\beta]\rangle=0\,.
\end{equation}
Given these assumptions, it follows that
\begin{widetext}
\begin{equation}
C_{11}=
\left(
\begin{array}{cccccc}
2\sigma_p^2 + \sigma_n^2 & \sigma_p^2 & 0 & 0 & 0 & 0 \\
\sigma_p^2 & 2\sigma_p^2 + \sigma_n^2 & 0 & 0 & 0 & 0 \\
0 & 0 & 2\sigma_p^2 + \sigma_n^2 & \sigma_p^2 & 0 & 0 \\
0 & 0 & \sigma_p^2 & 2\sigma_p^2 + \sigma_n^2 & 0 & 0 \\
0 & 0 & 0 & 0 & 2\sigma_p^2 + \sigma_n^2 & \sigma_p^2 \\
0 & 0 & 0 & 0 & \sigma_p^2 & 2\sigma_p^2 + \sigma_n^2 \\
\end{array}
\right)\,,
\end{equation}
\end{widetext}
\begin{eqnarray}
&&C_{12}=
\left(
\begin{array}{cccccc}
0 & 0 & 0 & 0 & 0 & 0 \\
0 & 0 & 0 & 0 & 0 & 0 \\
0 & \sigma_p^2 & 0 & 0 & 0 & -\sigma_p^2 \\
0 & 0 & 0 & 0 & 0 & -\sigma_p^2 \\
0 & 0 & -\sigma_p^2 & \sigma_p^2 & 0 & 0 \\
0 & 0 & -\sigma_p^2 & 0 & 0 & 0 \\
\end{array}
\right)\,,
\\[5pt]
&&C_{13}=
\left(
\begin{array}{cccccc}
0 & 0 & 0 & 0 & -\sigma_p^2 & 0 \\
0 & 0 & 0 & 0 & -\sigma_p^2 & 0 \\
0 & 0 & 0 & 0 & 0 & 0 \\
0 & 0 & 0 & 0 & 0 & 0 \\
0 & -\sigma_p^2 & 0 & 0 & 0 & 0 \\
\sigma_p^2 & -\sigma_p^2 & 0 & 0 & 0 & 0 \\
\end{array}
\right)\,,
\\[5pt]
&&C_{14}=
\left(
\begin{array}{cccccc}
0 & 0 & 0 & -\sigma_p^2 & 0 & 0 \\
0 & 0 & 0 & -\sigma_p^2 & 0 & 0 \\
-\sigma_p^2 & 0 & 0 & 0 & 0 & 0 \\
-\sigma_p^2 & 0 & 0 & 0 & 0 & 0 \\
0 & 0 & 0 & 0 & 0 & 0 \\
0 & 0 & 0 & 0 & 0 & 0 \\
\end{array}
\right)\,,
\\[5pt]
&&C_{15}=
\left(
\begin{array}{cccccc}
0 & 0 & 0 & 0 & 0 & 0 \\
0 & 0 & 0 & 0 & 0 & 0 \\
0 & 0 & 0 & 0 & 0 & 0 \\
0 & 0 & 0 & 0 & 0 & 0 \\
0 & 0 & 0 & 0 & 0 & 0 \\
0 & 0 & 0 & 0 & 0 & 0 \\
\end{array}
\right)\,.
\end{eqnarray}
In terms of these sub-matrices, the full covariance matrix is
\begin{equation}
C=
\left(
\begin{array}{ccccc}
C_{11} & C_{12} & C_{13} & C_{14} & C_{15} \\
C_{12}{}^T & C_{11} & C_{12} & C_{13} & C_{14} \\
C_{13}{}^T & C_{12}{}^T & C_{11} & C_{12} & C_{13} \\
C_{14}{}^T & C_{13}{}^T & C_{12}{}^T & C_{11} & C_{15} \\
C_{15}{}^T & C_{14}{}^T & C_{13}{}^T & C_{12}{}^T & C_{11} \\
\end{array}
\right)\,,
\end{equation}
and is shown schematically in Fig.~\ref{fig:cov_schematic}.

It is now an exercise in linear algebra to compute the eigenvectors
and eigenvalues of $C$. We use {\em MAPLE} \cite{Maple} to do this
calculation for us. There are 22 distinct eigenvalues, the smallest
of which is independent of the laser frequency noise. This minimal
eigenvalue
\begin{equation}
\lambda_{\rm min}\equiv\sigma_n^2
\end{equation}
is 9-fold degenerate. The nine eigenvectors corresponding to
$\lambda_{\rm min}$ orthogonally span a 9-dimensional vector
subspace of the 30-dimensional vector space of data points, elements
of which do not depend on the laser frequency noise (see
Table~\ref{tab:eigenvectors}). Thus, these eigencombinations
correspond to the TDI-like variables for this particular example. In
principle one could compute the covariance matrix for the entire set
of LISA data ($\sim 10^9$ data points), though in practice this
should not be necessary.  In our example we need only four time
stamps to fully characterise the covariance matrix, and a long time
sequence of data from the example can be generated using the same
set of eigencombinations of Doppler measurements. We will refer to
the time series generated by these combinations as an
\emph{eigenstreams}. One can show e.g., that the Sagnac combination
for spacecraft 1 (denoted $\alpha(t)$ in \cite{livrev}) is a linear
combination of eigenvectors: $-{\bf e}_4-{\bf e}_7+{\bf e}_8$.
Explicitly,
\begin{equation}
\alpha[5]\equiv s'_1[5]+s'_3[3]+s'_2[2]-s_1[5]-s_2[2]-s_3[1]\,,
\end{equation}
which is just the discretised-version of
\begin{align}
\alpha(t)\equiv&
\left[s'_1(t)+{\cal D}_2s'_3(t)+{\cal D}_1{\cal D}_2s'_2(t)\right]
\nonumber
\\
&-\left[s_1(t)+{\cal D}_3s_2(t)+{\cal D}_1{\cal D}_3s_3(t)\right]
\\
=&\left[s'_1(t)+s'_3(t-L_2)+s'_2(t-L_1-L_2)\right]
\nonumber
\\
&-\left[s_1(t)+s_2(t-L_3)+s_3(t-L_1-L_3)\right]\,,
\end{align}
(cf.\ Eq.~(42) in Ref.~\cite{livrev}) appropriate for the arm
lengths given in Eq.~(\ref{e:armlengths}).

\begin{table*}
\begin{ruledtabular}
\begin{tabular}{cccccccccc}

Data & ${\bf e}_1$ & ${\bf e}_2$ & ${\bf e}_3$ & ${\bf e}_4$ &
${\bf e}_5$ & ${\bf e}_6$ & ${\bf e}_7$ & ${\bf e}_8$ & ${\bf e}_9$ \\

\hline

$s_1[1]$  & 0 & 0 & 0 & 0 & 0 & 0 & 0 & 0 & 0 \\
$s'_1[1]$ & 0 & 0 & 0 & 0 & 0 & 0 & 0 & 0 & 0 \\
$s_2[1]$  & 0 & 0 & 0 & 0 & 0 & 1 & 0 & 0 & 0 \\
$s'_2[1]$ & 0 & 0 & 0 & 0 & 0 & 0 & 0 & 0 & 0 \\
$s_3[1]$  & 0 & 0 & 0 & 1 & 0 & 0 & 0 & 0 & 0 \\
$s'_3[1]$ & 0 & 1 & 0 & 0 & 0 & 0 & 0 & 0 & 0 \\
$s_1[2]$  & 0 & 0 & 0 & 0 & 0 & 0 & 0 & 0 & 0 \\
$s'_1[2]$ & 0 & 0 & 0 & 0 & 0 & -1 & 0 & 0 & 0 \\
$s_2[2]$  & 0 & 0 & 0 & 0 & 0 & 0 & 1 & 0 & 0 \\
$s'_2[2]$ & 0 & 0 & 0 & -1 & 0 & 0 & 0 & 0 & 0 \\
$s_3[2]$  & 1 & 0 & 0 & 0 & 0 & -1 & 0 & 0 & 0 \\
$s'_3[2]$ & 0 & 0 & 0 & 0 & 0 & 0 & 0 & 0 & 1 \\
$s_1[3]$  & 0 & -1 & 0 & 0 & 0 & 0 & 0 & 0 & 0 \\
$s'_1[3]$ & 0 & 1 & 0 & 1 & 0 & 0 & -1 & 0 & 0 \\
$s_2[3]$  & 1 & 0 & 0 & 0 & 0 & 0 & 0 & 0 & 0 \\
$s'_2[3]$ & -1 & 0 & 0 & 0 & 0 & 1 & 0 & 0 & 0 \\
$s_3[3]$  & 0 & 0 & 1 & 0 & 0 & 0 & 0 & 0 & 0 \\
$s'_3[3]$ & 0 & 0 & 0 & 0 & 0 & 0 & 0 & 1 & 0 \\
$s_1[4]$  & 0 & 0 & 0 & 0 & 0 & 1 & 0 & 0 & -1 \\
$s'_1[4]$ & 0 & 0 & 0 & 0 & 0 & -1 & 0 & 0 & 1 \\
$s_2[4]$  & 0 & 0 & 1 & -1 & 0 & 0 & 1 & 0 & 0 \\
$s'_2[4]$ & 0 & 0 & -1 & 0 & 0 & 0 & 0 & 0 & 0 \\
$s_3[4]$  & 0 & 0 & 0 & 0 & 1 & 0 & 0 & 0 & 0 \\
$s'_3[4]$ & 0 & 0 & 0 & 0 & 0 & 1 & 0 & 0 & 0 \\
$s_1[5]$  & 0 & 0 & 0 & -1 & 0 & 0 & 1 & -1 & 0 \\
$s'_1[5]$ & 0 & 0 & 0 & 1 & 0 & 0 & -1 & 1 & 0 \\
$s_2[5]$  & 0 & 0 & 0 & 0 & 1 & 1 & 0 & 0 & 0 \\
$s'_2[5]$ & 0 & 0 & 0 & 0 & -1 & -1 & 0 & 0 & 0 \\
$s_3[5]$  & 0 & 0 & 0 & 1 & 0 & 0 & -1 & 0 & 0 \\
$s'_3[5]$ & 0 & 0 & 0 & -1 & 0 & 0 & 1 & 0 & 0 \\

\end{tabular}
\end{ruledtabular}
\caption{Eigenvectors corresponding to the 9-fold degenerate
eigenvalue $\lambda_{\rm min}\equiv\sigma_n^2$.}
\label{tab:eigenvectors}
\end{table*}


\section{Discussion}

Although the LISA model considered above is greatly simplified,
only considering fixed arm-lengths related by small integers and a
basic noise model, we believe that the principal component
approach is suitable for more sophisticated LISA models.  In
particular, we have indicated that the likelihood is the natural
generating function for orthogonal LISA data streams that are not
dominated by laser frequency noise. This reduced set of
eigenstreams is a sufficient basis (and possibly more than
sufficient) to carry out all possible astronomy with LISA. The
remaining eigenstreams contain information on the laser stability
and are therefore important for instrumental diagnostics rather
than astrophysics.

As we have emphasised, the frequency-noise-free eigenstreams are
directly equivalent to TDI variables, but emerge naturally as a
direct consequence of the likelihood analysis and so have a
meaning that is directly relevant to subsequent data analysis. For
example, if we take a model $M$ of how an astrophysical source, or
a number of sources, would appear in the LISA data and assume this
model depends on a set of parameters $\mathbf{a}$ (such as sky
position, polarisation angle, etc.), then the joint posterior
probability of these parameters is simply
\begin{equation}
p(\mathbf{a}|\{\mathbf{e_i}\},M) \propto p(\mathbf{a}|M)\, \prod_i
p(\mathbf{e}_i|\mathbf{a},M)
\end{equation}
where $\mathbf{e_i}$ are the orthogonal frequency-noise-free
eigenstreams generated from the data.  In this way the LISA data
analysis problem is cast in the powerful framework of classic
inference, suitable for attack by standard search and exploration
algorithms including Markov Chain Monte Carlo methods (e.g.
\cite{Umstat1, Umstat2, Cornish}).

More complex models of the LISA spacecraft will of course increase
the complexity and size of the covariance matrix, but the basic
principle will remain the same. For example, differing laser shot
noise in the arms will break the eigenvalue degeneracy in the above
example. In addition, the lengths $L_i$ and $L_i'$ cannot be assumed
equal when there is relative spacecraft motion and as a result we
expect the larger covariance matrix to yield 2nd generation TDI
variables from its eigenstreams.

The eigenstreams are defined in terms of their minimal variance
rather than the amount of astronomy they contain, and there is no
reason why they should all contain astronomical information. The
eigenstreams, or combinations of eigenstreams, that are devoid of
astronomical data are `zero-signal solutions' \cite{tintolarson},
and can help in instrument diagnostics in circumstances when the
astronomical signals would dominate (such as in the low frequency
confusion limit of LISA) .

We are reminded here of the importance of careful data acquisition
and sampling for LISA. It is well-understood that only identically
represented samples of the laser frequency noise will cancel
effectively, and there could be times when the covariance between
Doppler channels from this noise is reduced. This would be encoded
as an increase in the baseline-dependent noises terms, $n_i$ and
$n_i'$, and in the limit  as $\sigma_p^2$ approached $\sigma_n^2$
the eigenvalues of the covariance matrix would cease to break into
two groups.

Finally, although PCA is often used for data compression and, as we
have shown, the TDI-like eigenstreams contain all the `good quality'
astronomical data from LISA, we are \emph{not} proposing that these
be generated on the spacecraft and relayed back to Earth. The saving
in telemetry bandwidth would be clear, but having the raw Doppler
data on Earth would greatly enhance the flexibility of the analysis
and the robustness of the mission.



\end{document}